\documentclass[numsec,webpdf,modern,medium,namedate]{oup-authoring-template}

\onecolumn

\graphicspath{{Fig/}}

\usepackage{tabulary}
\usepackage{tabularx}
\usepackage{float}
\usepackage{graphicx}
\usepackage{makecell}
\usepackage{booktabs}
\usepackage{caption}
\usepackage{longtable}
\usepackage{subcaption}
\usepackage{amsmath}
\usepackage{lscape}
\usepackage{natbib}
\onecolumn % for one column layouts

\graphicspath{{Fig/}}

\theoremstyle{thmstyleone}%
\theoremstyle{thmstyletwo}%
\theoremstyle{thmstylethree}%

\begin{document}

\journaltitle{Prepared for submission to Journals of the Royal Statistical Society Series A}
\DOI{DOI HERE}
\copyrightyear{XXXX}
\pubyear{XXXX}
\access{Advance Access Publication Date: Day Month Year}
\appnotes{Original article}

\firstpage{1}

%\subtitle{Subject Section}

\title[Evaluating MRP with Spatial Priors]{Evaluating Multilevel Regression and Poststratification with Spatial Priors with a Big Data Behavioural Survey}

% Ensure that you provide the full postal address of each affiliation, including the country name and, if available, the email address of each author.
\author[1,$\ast$]{Aja Sutton \ORCID{0000-0002-0579-8301}} 
\author[2,3]{Zack W. Almquist \ORCID{0000-0002-1967-123X}}  %\ORCID{}
\author[3,4]{Jon Wakefield} %\ORCID{}

\authormark{Sutton et al.}

\address[1]{\orgdiv{Department of Environmental Social Sciences}, \orgname{Stanford University}, \orgaddress{\street{Yang \& Yamazaki (Y2E2) Building, 473 Via Ortega, Room 350}, \postcode{94305}, \state{California}, \country{United States of America}}}
\address[2]{\orgdiv{Department of Sociology}, \orgname{University of Washington}, \orgaddress{\street{Savery Hall, Room 211, 4100 Spokane Ln., Seattle}, \postcode{98195}, \state{Washington}, \country{United States of America}}}
\address[3]{\orgdiv{Department of Statistics}, \orgname{University of Washington}, \orgaddress{\street{C-138 Padelford Hall, Box 35435, Seattle}, \postcode{98195}, \state{Washington}, \country{United States of America}}}
\address[4]{\orgdiv{Department of Biostatistics}, \orgname{University of Washington}, \orgaddress{\street{Hans Rosling Center for Population Health, 3980 15th Avenue NE, Box 351617, Seattle}, \postcode{98195}, \state{Washington}, \country{United States of America}}}

\corresp[$\ast$]{Address for correspondence. Aja Sutton, Population Research Center, Portland State University, 506 SW Mill St., Suite 780, Portland, OR, 97201, USA. \href{Email:ajas@pdx.edu}{ajas@pdx.edu}} 

\received{Date}{0}{Year}
\revised{Date}{0}{Year}
\accepted{Date}{0}{Year}

%\editor{Associate Editor: Name}

%\abstract{
%\textbf{Motivation:} .\\
%\textbf{Results:} .\\
%\textbf{Availability:} .\\
%\textbf{Contact:} \href{name@email.com}{name@email.com}\\
%\textbf{Supplementary information:} Supplementary data are available at \textit{Journal Name}
%online.}

%maximum length of 200 words: 191 below
\abstract{Multilevel regression and poststratification (MRP) is a computationally efficient estimation method that can quickly produce improved population-adjusted estimates with limited data. Recent computational advancements allow efficient, relatively simple, and quick approximate Bayesian estimation for MRP. As population health outcomes of interest including vaccination uptake are known to have spatial structure, precision may be gained by including space in the model. We test a spatial MRP model that includes terms that smooth across demographics and geographic areas using a large, unrepresentative survey. We produce California county-level estimates of first-dose COVID-19 vaccination up to June 2021 using classic and spatial MRP models, and poststratify using data from the US Census Bureau's American Community Survey. We assess validity using reported first-dose vaccination counts from the Centers for Disease Control (CDC). Neither classic nor spatial MRP models performed well, highlighting: spatial MRP may be most appropriate for richer data contexts, some demographics in the survey data are over-sampled and -aggregated, producing model over-smoothing, and a need for survey producers to share user-representative metrics that allow better benchmarking of estimates.}

%keywords/phrases
\boxedtext{
\begin{itemize}
\item Bayesian spatial statistics
\item multilevel regression and poststratification
\item survey methods
\end{itemize}}

\maketitle
\section{Introduction} 
Multi-level regression and poststratification \citep{GelmanLittle1997,Little1993,ParkEtAl2004,WangEtAl2015,Valliant2020} (``classic MRP'') is a computationally efficient estimation method. It is used to smooth and adjust for non-representativeness in survey samples, and can also be used as a small area estimation (SAE) method to produce subnational areal estimates where there are limited or no specific survey data available \citep{Lopez-MartinEtAl2022}. It uses partial pooling to make predictions of unknown area-level estimates using total subpopulation trends \citep{BafumiGelman2006}, irrespective of the spatial relationships of subnational areal units. This may be concerning because many population health outcomes of interest, including disease spread or vaccination uptake, are known to have spatial structure (clear relationships across geographic space) \citep{WallerCarlin2010}. Including spatial structure in the model can improve precision, though excluding space does not necessarily lead to bias. Yet, it also is well-known that health outcomes often have demographic structure and subpopulations from neighbouring places may share similarities in health and health behaviour \citep[for example, see:][]{DiezRoux2001,DiezRouxEtAl2001}. 
    
Recently, \citet{GaoEtAl2021} proposed a ``spatial MRP" in which they include in the classic MRP model a spatial random effect on area-level administrative units that defines its structural relationship through a first-order spatial contiguity (neighbourhood) matrix and employs a modified Besag-York-Mollié (BYM2) model \citep{RieblerEtAl2016,MorrisEtAl2019}. \textit{In silico}, this was found to reduce absolute bias among most area-level population count estimates in a model including demographics (i.e., education and race/ethnicity) by approximately ten percent in comparison to classic MRP. While simulation experiments have demonstrated the efficacy of spatial MRP, it has not been established in the demography, survey, and social science literature as to when it can be reliably employed for improved SAE and when other methods should be considered. 
    
Typically, spatial Bayesian methods have required computationally demanding Markov Chain Monte Carlo (MCMC) approaches that are slow to implement, especially for spatial models which are by their nature extremely complex. Now, advancements in computation using integrated nested Laplace approximations (INLA) implemented through the \texttt{r-INLA} package \citep[see: www.r-inla.org,][]{SimpsonEtAl2017} allow us to approximate Bayesian inference for complex hierarchical models with and without spatial smoothing, produce estimates, and report uncertainty extremely quickly and conveniently without MCMC. In this article, we compare classic MRP and spatial MRP -- in which we include a BYM2 spatial term that smooths \textit{along or within demographic categories} and\textit{ between areas} -- using INLA, and a large, unrepresentative survey: the Delphi Group at Carnegie Mellon University’s United States-specific COVID-19 Trends and Impact Survey (CTIS), administered in partnership with Facebook \citep{SalomonEtAl2021}. The CTIS provides an exciting opportunity to test spatial MRP with real-world data: \citet{BradleyEtAl2021} have shown that the CTIS overestimates state-level vaccination counts by a substantial margin compared to baseline counts reported by the Centers for Disease Control and Prevention (CDC). We compare county- and aggregated state-level estimates with CDC baseline estimates, and discuss benefits and drawbacks of classic and spatial MRP, broadly.

This paper is laid out as follows: First, we discuss the specifics of the datasets used to model and validate the results (the CTIS and CDC estimates), including descriptive statistics contained in the CTIS sample. We also describe the poststratification data used (taken from US Census Bureau's American Community survey), including the manner in which we harmonized these with the CTIS to allow for pooling. Second, we describe four Bayesian hierarchical models (two that do not account for spatial dependence, and two with spatial smoothing). We delineate the poststratification process, including the calculations used to arrive at an aggregate estimate of first-dose vaccination, and also briefly describe the spatial and non-spatial models we build and compare in our analysis. Third, we report the results of the comparison of the modelled estimates with CTIS mean survey estimates and the CDC baseline estimates at the county level, as well as county-level estimates aggregated to the state level. Finally, we discuss the implications of these results and summarize these in the conclusion.

\section{Methods} \label{sec2}

This paper compares classic and spatial MRP methods by building several Bayesian hierarchical models using the United States' specific version of the CTIS. The outcome of interest is the number of first-dose COVID-19 vaccinations as of June 30, 2021 in California counties. Demographic strata were harmonized between this survey and the poststratification data prior to modelling. To produce estimates, we modelled the probability of first-dose vaccination against COVID-19 as of June 30, 2021 at the county level by demographic strata (age and educational attainment) using the US-specific version of the CTIS. Male and female models were built separately. To produce population estimates from these, we poststratified using 2017-2021 5-year American Community Survey population estimates with the harmonized demographic strata. To assess performance, we compared model estimates, CDC reported counts, and CTIS mean survey estimates of first-dose vaccination as of June 30, 2021. Previous research suggests COVID-19 vaccine hesitancy in the US has been driven by differences in experiences of race/ethnicity, where members of historically marginalized groups were less likely to receive vaccines \citep[e.g.,][]{NguyenEtAl2022,KhubchandaniMacias2021,WillisEtAl2021}. Though this trend initially held during the COVID-19 pandemic in California, there is evidence that by early 2021 COVID-19 vaccine trends were primarily driven by educational attainment: higher education was associated with greater likelihood of vaccination \citep{ThomasEtAl2021}. Trends in the CDC California first-dose vaccination estimates also suggest vaccination decisions varied by age, educational attainment, and county (race/ethnicity data were not provided) \citep{CentersforDiseaseControlandPrevention2023}. In the absence of race/ethnicity baseline data, we focus on variables available in both the CTIS and CDC: sex, age, and educational attainment. Detailed descriptions of these data and methods, including the generalized modelling approach, are laid out below.

\subsection{Data} \label{subsec1}
\subsubsection{US-specific CTIS} \label{subsec1.1}
This paper employs a novel dataset collected by the Delphi Group at Carnegie Mellon University in partnership with Facebook to establish a large US COVID-19 Trends and Impact Survey (CTIS) \citep{SalomonEtAl2021}. The United States' specific portion of the CTIS is an individual-level survey conducted from April 2020 to June 2022, and administered daily to approximately 40,000 residents of the United States (1-2\% response rate; 10-20\% completion rate). It includes respondents’ demographic and relative geographic information (e.g., ZIP codes and county FIPS codes). It focuses on respondents’ experiences and effects of the pandemic, including COVID-19-like symptoms, respondents’ behaviours (including mask-wearing and social distancing), and impacts on their mental health, economic situation, and overall health. \citet{BradleyEtAl2021} identified issues with the CTIS' survey weights: for vaccination counts in California, they are non-representative at geographies smaller than state-level, providing an opportunity to test spatial MRP for county-level estimates using these data.

\subsubsection{First-dose COVID-19 vaccination counts} \label{subsec1.2}
First-dose vaccination counts recorded by the Centers for Disease Control (CDC) were taken as the baseline for comparison by age and sex for model validation at the state level \citep{CentersforDiseaseControlandPrevention2023a}. The CDC total first-dose vaccination counts were used for validation at the county-level (i.e., data at the county level were not available by sex and age and represent total population counts) \citep{CentersforDiseaseControlandPrevention2023}. For this analysis, CTIS data were limited to California residents who began the survey in June 2021. This was done to obtain the same probability of receiving a first dose of COVID-19 vaccination as the CDC-reported first-dose vaccination counts in the state of California as of June 30, 2021; and also the same probability of said outcome within county-level total first-dose vaccination count data from the CDC on the same date. There were 49,700 individual CTIS respondents over the age of 18 years for whom information regarding receipt of first COVID-19 vaccination dose, sex, age, educational attainment, and county FIPS code were available; 57 of a total 58 counties were represented in these data.

\subsubsection{Descriptive statistics of CTIS and poststratification weights} \label{subsec1.3}
Descriptive statistics for the CTIS data are reported in Table \ref{tab:tab1_descriptivestatsctis} (by age and sex; see below) and Table \ref{tab:ctis_direct_agesexedu} (by age, sex, and educational attainment; see: Appendix). CDC-reported first-dose vaccination counts for the state of California as of June 30, 2021 by sex and age are reported in Table \ref{tab:cdc_counts_bysexage} (for this and the total first-dose vaccination counts at the county level, Table \ref{tab:cdc_counts_bysexage_county}, see the Appendix). 
 \vspace{10pt}
\begin{table}[H]
    \centering
    \footnotesize
    \caption{Mean survey estimates of vaccination status in the COVID-19 Trends and Symptoms Survey (CTIS) as of June 30, 2021, by sex and age}
    \label{tab:tab1_descriptivestatsctis}
    \begin{tabular}{@{}llcc@{}}
        \toprule
        \textbf{Sex} & \textbf{Age} & \textbf{\makecell{Mean}} & \textbf{95\% Confidence Interval} \\ 
        \midrule
        Female& 18-24 years & 0.54& (0.51, 0.57)\\ 
        & 25-64 years & 0.53 & (0.52, 0.53) \\ 
        & 65 years and over & 0.55& (0.53, 0.57) \\ 
        \midrule
        Male& 18-24 years & 0.46& (0.43, 0.49)\\ 
        & 25-64 years & 0.47 & (0.47, 0.47)\\ 
        & 65 years and over & 0.45& (0.43, 0.47) \\ 
        \bottomrule
    \end{tabular}

\end{table}
\vspace{10pt}

Poststratification data were produced using the U.S. Census Bureau’s 2017-2021 5-year American Community Survey (ACS) estimates \citep{U.S.CensusBureau2022} (Table B15001), representing estimated population counts by age, sex, and education at the county level. The demographic strata described in Tables \ref{tab:tab1_descriptivestatsctis} and \ref{tab:ctis_direct_agesexedu} were harmonized to match those in the poststratification data. As such, possible poststratification variables for California counties included sex (male, Female), age (18-24 years, 25-64 years, 65 years and over), and educational attainment (less than high school, high school or equivalent, some college, associate’s degree, bachelor’s degree, professional or graduate degree). For simplicity, in this analysis zeroes in the ACS data were treated as ``true” zeroes.
  
\subsection{Modelling Approach: Spatial MRP} \label{subsection2}
With adequate data, it would be possible to take the empirical proportion of the cross-classification of age and education by sex and first-dose vaccination status. In the absence of these ideal circumstances, we instead aim to model the number of people aged 18 years and older who had received their first COVID-19 vaccination (hereafter, ``first-dose vaccination”) in each county in the state of California as of June 30, 2021.  We can consider the outcome of interest in this context as a binary variable (having received first-dose vaccination, or not), where we are primarily interested in the proportion of a county’s total number of vaccinated individuals (these can later be aggregated to the state level for further comparison). Under these assumptions, it is natural to specify a binomial model \citep{ParkEtAl2004}. In our models, we assign a Beta-binomial distribution to the outcome to handle within-strata variability between counties and attributed to the unrepresentative nature of the survey sample \citep[see e.g.,][]{DongWakefield2021}. We describe several spatial and non-spatial Bayesian hierarchical models to estimate the posterior marginal probability of first-dose vaccination across several demographic groups by age, education, and sex (both sexes are modelled separately). An overview of all models produced in this analysis is provided in Table \ref{tab:tab2_modelslist}. Once the models are fit via \texttt{r-INLA} \citep{RueEtAl2009,MartinsEtAl2013}, we poststratify using the process also described in Section \ref{subsection2.4}.
    
\subsubsection{Model notation} \label{subsection2.1}
Define $n_{ijk}$ as the population in age stratum $j = \{1,2,J = 3\}$ (1 = 18-24 years, 2 = 25-64 years, and 3 = 65 years and over; the reason for the limited number of strata is described in section \ref{subsec1.3}), and educational attainment stratum (henceforth, ``education stratum") $k = \{1,..., K = 6\}$ (where 1 = less than high school, 2 = high school or equivalent, 3 = some college, 4 = associate’s degree, 5 = bachelor’s degree, and 6 = professional or graduate degree) within county $i = \{1,...,I = 58\}$. Recall we are modelling male and female separately. Therefore, for each sex the total population in each county is $n_i = \sum^{J}_{j=1}\sum^{K}_{k=1} n_{ijk}$, and the total population of California is $N = \sum^{I}_{i=1} \sum^{J}_{j=1} \sum^{K}_{k=1} n_{ijk}$. With respect to each sex, let $y_{ijk}$ be the number of people who received a first-dose vaccination out of the total $n_{ijk}$ people in county $i$, age stratum $j$, and education stratum $k$. 

\subsubsection{Hierarchical models without spatial dependence} \label{subsection2.2}
In classic MRP, geographic variation is accounted for by modelling the outcome within each independent geographic area  -- e.g., by assuming counties are independent and identically distributed (IID). To reflect this, we specify a non-spatial hierarchical model with $\alpha$ the intercept, fixed effects by age, $\lambda_j$, education, $\beta_{k}$, and an IID random effect by county, $\epsilon_i$,
\begin{equation} \label{eq3.1}
    \begin{gathered} 
        y_{ijk}|\theta_{ijk} \sim BetaBinomial(n_{ijk},\theta_{ijk},d), \\
        logit(\theta_{ijk}) = \alpha + \lambda_j + \beta_{k} +  \epsilon_{i},\\
                   \epsilon_i  \overset{iid}{\sim}  Normal(0, \sigma_\epsilon^2).
    \end{gathered}
\end{equation}
with $d$ the overdispersion parameter for the beta-binomial distribution. A fixed effect is most appropriate for age because of data sparseness (i.e., too few age strata).

It is also possible to include smoothing across demographic strata. We can specify $\beta_k$ as a first-order random effect (RW1) by education -- a form of computationally convenient Gaussian Markov Random Field (GMRF) model \citep{RueHeld2005}. In this case, a RW1 random effect smooths across the adjacent neighbors. We specify a hierarchical model,
\begin{equation} \label{eq4}
         \begin{gathered} 
                        y_{ijk}|\theta_{ijk} \sim BetaBinomial(n_{ijk}, \theta_{ijk}, d), \\
                        logit(\theta_{ijk}) = \alpha + \lambda_j + \beta_{k} + \epsilon_{i}, \\
                        \underline{\beta} = [\beta_1,...,\beta_{K}] \sim RW1(\sigma^2_\beta), \\
                        \epsilon_i \overset{iid}{\sim} Normal(0, \sigma_\epsilon^2),
        \end{gathered}
    \end{equation}
with a fixed effect on age $\lambda_j$, and IID by county random effects $\epsilon_i$. To include the intercept, $\alpha$, we include a sum-to-zero constraint on $\beta$ for identifiability required by random walk models \citep{BesagEtAl1991}. Additionally, we use penalized complexity (PC) prior specifications for the hyperpriors \citep{SimpsonEtAl2017}. PC priors penalize deviance from a null model (in this case, one without the random effects). PC prior choices are defined by a probability statement, $P(\sigma > u) = a$, where the user specifies $u$ and $a$. Here, we specify $u = 0.5$ and $a =0.1$.

\subsubsection{Spatial models} \label{subsection2.3}
 As vaccination trends are known to have varied across Californian counties according to CDC estimates, the outcome may have spatial dependence, and some level of measurable spatial autocorrelation \citep{Banerjee2016}.  We describe two spatial models: one that smooths across counties based solely on an adjacency matrix of first-order (immediately geographically proximate) neighbouring counties, and one that smooths across counties' education strata.

It may be reasonable to expect subpopulations in neighbouring counties share some commonality with respect to vaccination uptake that has not been observed in the data (i.e., a BYM2 spatial random effect that smooths unobserved variance across neighbouring counties, $\gamma_i$, without it also explicitly smoothing across demographics). We may thus specify a spatial hierarchical model that includes a fixed effect for age, $\lambda_j$, RW1 by education, $\beta_k$, and also shares information between each neighbouring county using a BYM2 spatial random effect, $\gamma_i$ \citep{RieblerEtAl2016} :
\begin{equation} \label{eq2}
         \begin{gathered} 
                        y_{ijk}|\theta_{ijk} \sim BetaBinomial(n_{ijk}, \theta_{ijk}, d), \\
                        logit(\theta_{ijk}) = \alpha + \lambda_j + \beta_{k} + \gamma_i, \\
                        \underline{\beta} = [\beta_1,...,\beta_{K}] \sim RW1(\sigma^2_\beta), \\
                       \underline{\gamma_i} = [\gamma_{1},...,\gamma_{I}] \sim BYM2(\sigma^2_{\gamma},\phi),
        \end{gathered}
    \end{equation}
with $\alpha$ the intercept. In this case, the BYM2 allows for a joint prior on the unstructured and structured term. As with random walk models, BYM models including BYM2 require a sum-to-zero constraint for identifiability to include the intercept, $\alpha$ \citep{BesagEtAl1991}. Again, we use PC priors where those for the RW1 model are specified as above. With respect to the PC priors on the BYM2 model, for the overall precision parameter we set $u=1$ and $a=0.01$, corresponding to a 0.99 prior probability of having residual odds ratios smaller than 2. For the mixing parameter, we set $u=0.5$ and $a=2/3$ corresponding to a 67\% chance that more than 50\% of the total variation of the random effect has spatial structure. This represents a classic spatial Bayesian hierarchical model that aims to shrink unobserved between-county variance \citep[for details on the derivation of the PC prior for the BYM2 model, see Appendix 2 in][]{RieblerEtAl2016}.

Alternatively, it may be reasonable to expect subpopulations with similar demographics in neighbouring counties may share some commonality (variance) with respect to vaccination uptake observable across demographic strata. In this model, we smooth across counties within education group. Assuming individuals within the same education group from neighbouring counties share similar probabilities of first-dose vaccination, we specify a modified form of the Besag-York-Mollié (BYM) spatial random effect \citep{BesagEtAl1991,Besag1974}, a BYM2 spatial random effect that includes an additional IID random effect on county to smooth within and between counties across education, $\gamma_{ik}$:
\begin{equation} \label{eq1}
         \begin{gathered} 
                        y_{ijk}|\theta_{ijk} \sim BetaBinomial(n_{ijk}, \theta_{ijk}, d), \\
                        logit(\theta_{ijk}) = \alpha + \lambda_j + \gamma_{ik}, \\
                       \underline{\gamma_k} = [\gamma_{1k},...,\gamma_{Ik}] \sim BYM2(\sigma^2_{\gamma k},\phi_{k}).
        \end{gathered}
    \end{equation}
We include a fixed effect on age, $\lambda_j$, and use PC priors. This represents a complex spatial Bayesian hierarchical model that smooths between and within counties across education, whilst accounting for age. 

\subsubsection{Multilevel regression and poststratification} \label{subsection2.4}
To attempt to adjust for the non-representativeness of the CTIS data, we produce marginal posterior estimates of first-dose vaccination using weights built from the marginal estimates of each demographic stratum (``level") in each county, and also stratum-specific county population totals from census data. For each model, we sample one thousand draws of the marginal posterior probabilities $s$ of first-dose vaccination within each county $i$, age $j$, and educational attainment $k$ cell: $\theta^{(s)}_{ijk}$ where $s = \{1,...,S\}$ samples. From this, we can take the sample's median and 95\% credibility interval: $\theta^{(s)MED}_{ijk} (\theta^{(s)LO}_{ijk}, \theta^{(s)HI}_{ijk})$. 

The poststratification data are taken from the 5-year 2017-2021 ACS. These data represent the population counts for each demographic stratum by age ($J =$ 3 levels), and educational attainment ($K =$ 6 levels), for each county ($I =$ 58 counties): $J \times K \times I = 1,044$ rows in the poststratification table for each sex (2,088 total). Using the poststratification data, we calculate a poststratification weight, i.e., the proportion of the total population for each sex in county $i$, age $j$, and education $k$: $w_{ijk} = n_{ijk}/n_i$. Estimates of the proportion of the population at the state level can then be calculated: $\theta^{(s)} = \sum_{ijk} {\theta^{(s)}_{ijk} w_{ijk}}$. County-level population estimates are,    
\begin{equation} 
         \begin{gathered} 
    y^{(s)}_i = \sum_{j} \sum_{k} y^{(s)}_{ijk} = \sum_{j} \sum_{k} n_{ijk}\theta^{(s)}_{ijk}, 
           \end{gathered}
    \end{equation}
    and can be aggregated to the state level: $y^{(s)}= \sum_i y^{(s)}_i$. Total population estimates are produced by summing the male and female estimates at the desired geographic level (county or state).

\begin{table}[H]
    \footnotesize
    \centering
        \caption{A brief description and summary of specifications of the hierarchical logistic regression models tested in this analysis.}
        \label{tab:tab2_modelslist} 
    \begin{tabularx}{\textwidth}{@{}Xccc@{}}
        \toprule
        \textbf{Model Description} & \textbf{\makecell{Age\\Specification*}} & \textbf{\makecell{Educational\\Attainment\\Specification*}} & \textbf{\makecell{County\\Specification*}} \\ 
        \midrule
       
        fixed effect on age, fixed effect on education, IID by county &fixed&fixed&IID \\
        fixed effect on age, RW1 by education, IID by county&fixed&RW1&IID\\
        fixed effect on age, BYM2 by education&fixed&BYM2&IID(implicit)\\
        fixed effect on age, RW1 by education, BYM2 by county &fixed&RW1&BYM2\\
        \bottomrule
    \end{tabularx}
       \footnotesize
        \textit{*Key: IID = independent and identically distributed; RW1 = first-order random walk; BYM2 = modified Besag-York-Mollié (BYM2)\\
        ** The BYM2 model has an implicit IID by area (in this case, county).}
\end{table}
 
\section{Results}\label{sec3}

In this section, we report the results of the Bayesian hierarchical models and state- and county-level aggregated poststratified estimates of first-dose COVID-19 vaccination in California as of June 30, 2021. Bayesian hierarchical regression results for each model are reported in Table \ref{tab:inla_regression_results} in the Appendix. Table \ref{tab:lcporesults_bysex} provides a summary of the model selection results using the mean logarithmic conditional predictive ordinate (LCPO) \citep{Geisser1993,GneitingRaftery2007,HeldEtAl2010}. For both female and male models, there was an equal preference for both models without spatial dependence (i.e., those containing IID by county), as well as the fixed effect on age, RW1 by education, BYM2 by county models. A parsimonious approach to model selection would suggest the simplest model -- fixed effects on age and education, IID by county -- to be preferable for both sexes.

{\footnotesize
\begin{longtable}{@{}l c c @{}}
    \caption{Mean logarithmic conditional predictive ordinate (LCPO) for county-level, sex-specific models}
    \label{tab:lcporesults_bysex}\\
    \toprule
    \textbf{Model Name} & \textbf{Sex} & \textbf{LCPO} \\ 
    \midrule
    \endfirsthead 

    fixed effect on age, RW(1) by education, BYM2 by county & Female & 0.313 \\
    fixed effect on age, RW(1) by education, IID by county & Female & 0.313 \\
    fixed effect on age, fixed effect on education, IID by county & Female & 0.314 \\
    fixed effect on age, BYM2 by education & Female & 0.32 \\
    fixed effect on age, fixed effect on education, IID by county & Male & 0.333 \\
    fixed effect on age, RW(1) by education, BYM2 by county & Male & 0.333 \\
    fixed effect on age, RW(1) by education, IID by county & Male & 0.333 \\
    fixed effect on age, BYM2 by education & Male & 0.343 \\

\bottomrule
\end{longtable}
}

Results comparing modelled estimates aggregated from the county- to the state-level with CDC baseline estimates are visualized in Figure \ref{fig:1} and reported in Table \ref{tab:state_propestimates} in the Appendix. State-level aggregates for all four models captured the mean survey estimates produced from CTIS survey weights, but did not capture the CDC estimate for the proportion of individuals in California who had received the first-dose COVID-19 vaccination for June 30, 2021.

\begin{figure}[H]
\centering
     \includegraphics[width = 0.7\textwidth]{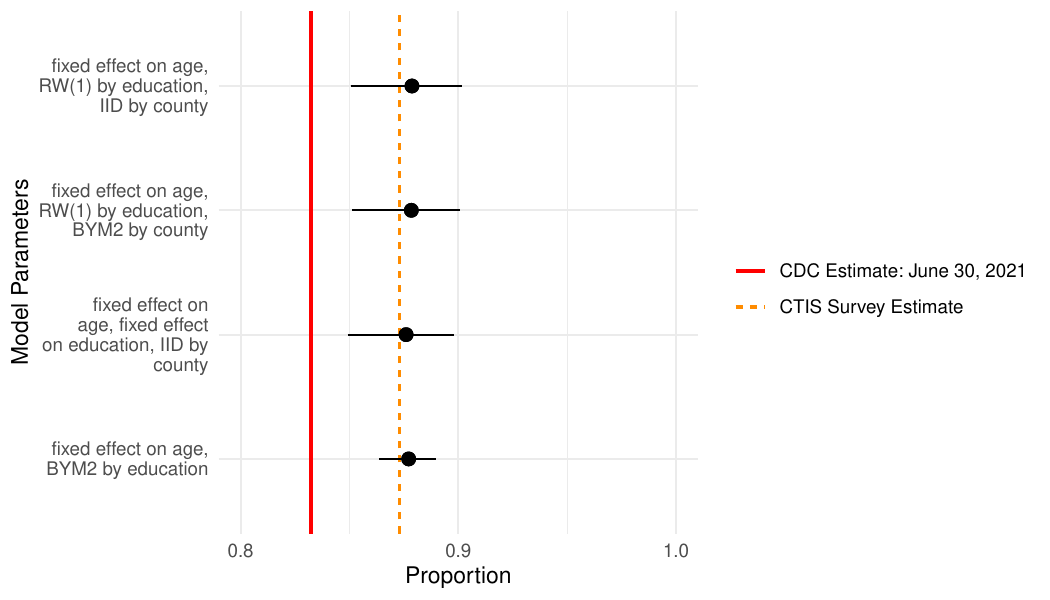}
     \caption{Results of the number of first-dose COVID-19 vaccinations modelled for both sexes at the county level and aggregated to the state level produced using classic MRP methods (IID by county), and spatial MRP methods (BYM2 models). Models were able to capture the mean survey estimate of first-dose COVID-19 vaccinations at the state level produced by the CTIS survey weights, but did not capture the CDC estimate for June 30, 2021.}
     \label{fig:1}
\end{figure}
\begin{figure}[H]
\centering
     \includegraphics[width=\textwidth]{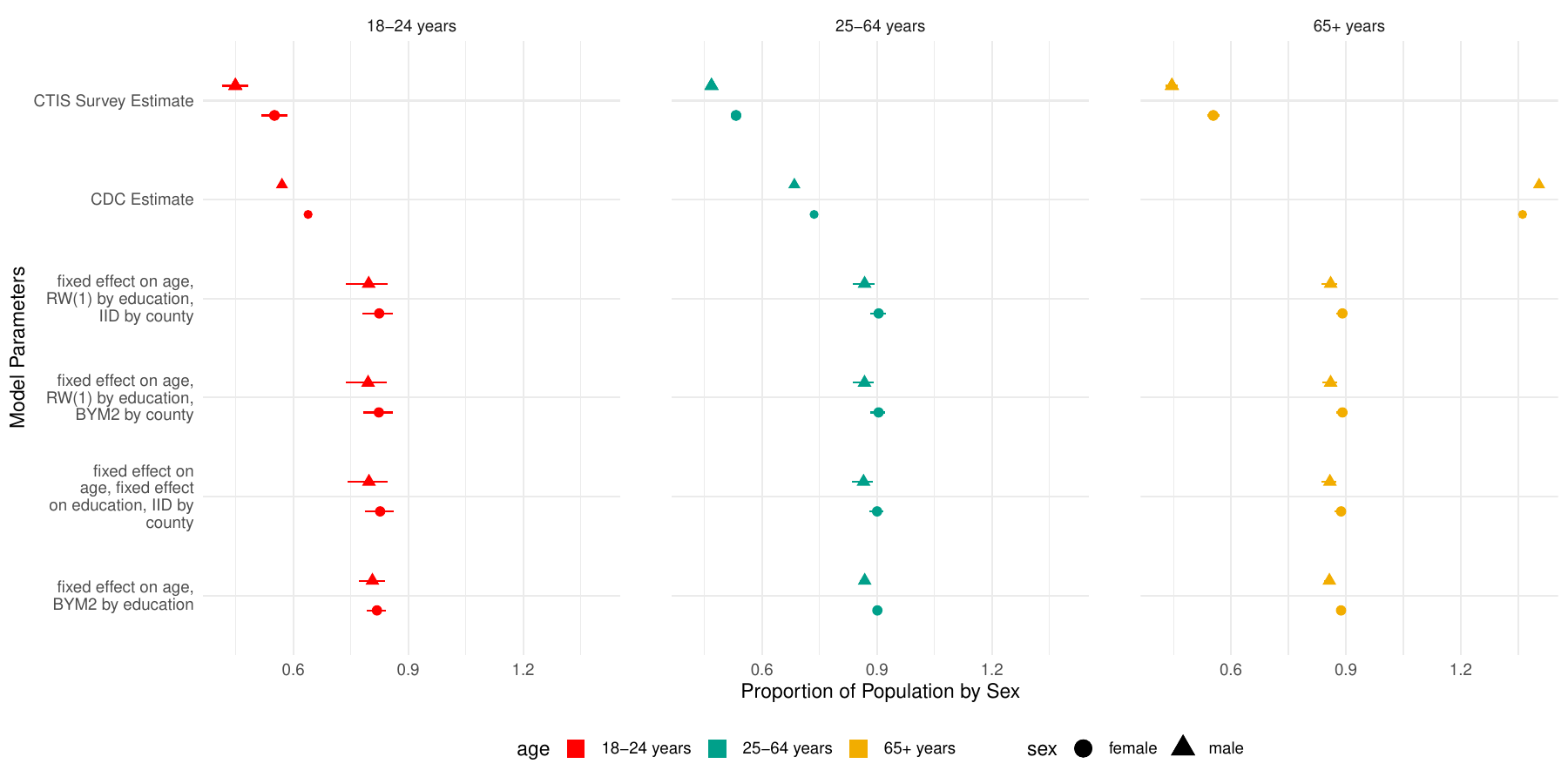}
     \caption{Results of the number of first-dose COVID-19 vaccinations modelled by sex and age aggregated to the state level produced using classic MRP methods (IID by county), and spatial MRP methods (BYM2 models).}
     \label{fig:1b}
\end{figure}
When we consider a finer grain visualization of estimates by age and sex (Figure \ref{fig:1b}), we see all models over-estimate vaccination for both sexes and all age groups. We also see an underlying issue with the CDC estimate for males and females aged 65 years and older is greater than the total population. We expect this is a result of either undocumented older individuals residing in California, or else because non-residents of California were vaccinated in the state. These results imply the CDC baseline estimate at the state level in Figure \ref{fig:1} is higher than the true proportion of Californians vaccinated on June 30, 2021.

\hfill

Results for the county-level estimates produced in this analysis are illustrated in Figure \ref{fig:2} and reported in Table \ref{tab:county_propestimates} in the Appendix. We recall the weights provided in the CTIS are known to be inappropriate for producing vaccination estimates below the state level. We use the county-level CDC estimates of first-dose vaccination for June 30, 2021 as the closest approximation of the truth with which to compare the validity of our modelled estimates. MRP estimates would be considered accurate if they captured CDC state-level estimates (and even more desirably, county-level estimates). In this case, spatial and classic MRP performed similarly and did not capture CDC estimates consistently, tending instead to overestimate vaccination prevalence; random effects models smooth with data they are provided, but are unable to handle this level of bias in the data.

\begin{figure}[H]
\centering
    \begin{subfigure}{\textwidth}
    \centering
    \includegraphics[width=0.99\textwidth]{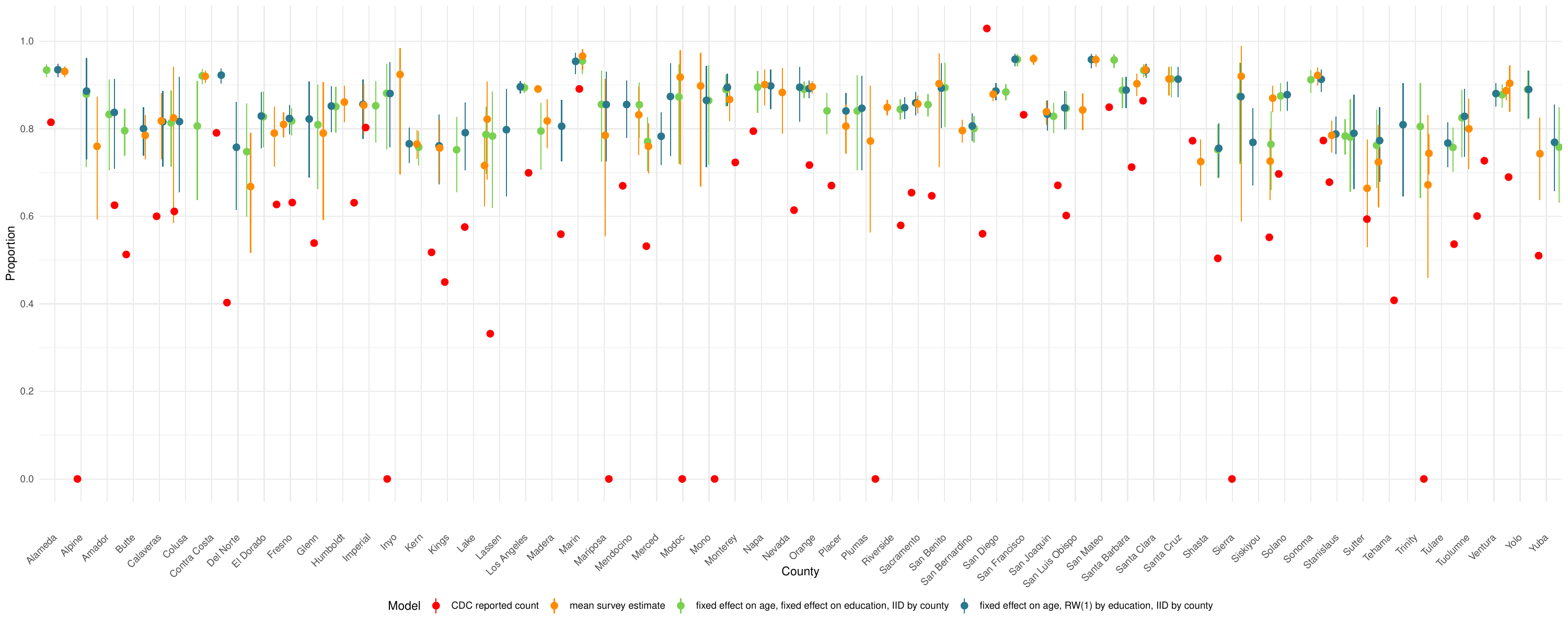}
    \caption{Classic MRP with non-spatial (IID) models.} \label{fig:2a}
    \end{subfigure}
    \\
    \begin{subfigure}{\textwidth}
    \centering
    \includegraphics[width=0.99\textwidth]{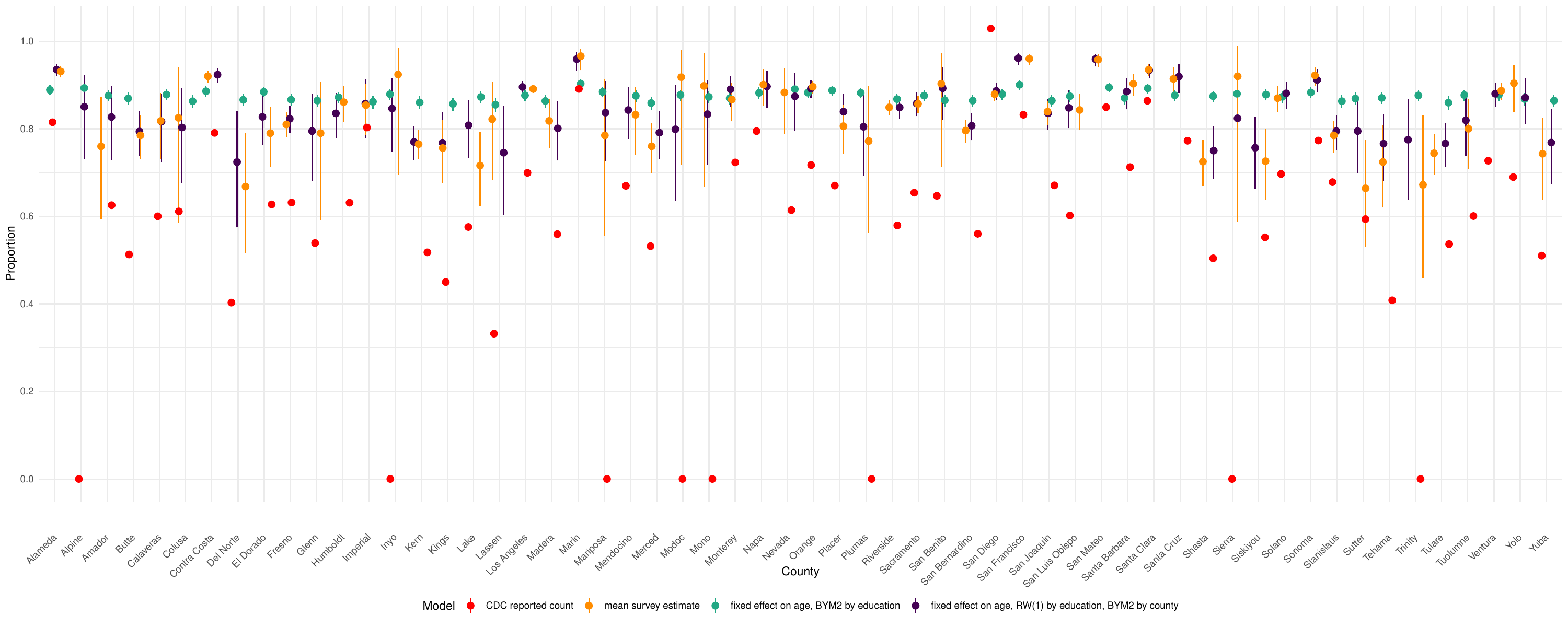}
    \caption{Spatial MRP with spatial (BYM2) models.} \label{fig:2b}
    \end{subfigure}
\caption{Results of the number of first-dose COVID-19 vaccinations modelled at the county level produced using (a) classic MRP methods (IID by county), and (b) spatial MRP methods (BYM2 models). In this case, MRP did not capture CDC estimates, or CTIS estimates at the county-level, tending instead to over-smooth.} \label{fig:2}
\end{figure}

\subsection{Limitations} \label{sec3.2}
The CTIS itself has a few limitations and errors. The Delphi Group describe these on the CTIS website \citep{DelphiGroup}. They acknowledge the survey weights are inappropriate for estimation below the state level. Additionally, between June 15 and July 4, 2021, a random 1\% of the sample was sent the instrument for the previous wave (10) rather than that which was current at the time (wave 11). The Delphi Group claim to have not detected trend discontinuities as a result of the error, despite differences in the two instruments. Additionally, between June 17 and 24, 2021, users using an Android phone to access the Facebook application through which the survey was administered were unable to open the app due to a bug in the in-app web browser. The Delphi Group reported that they had recovered 89.5\% of the initial decrease in total response volume as of June 26, 2021. Another limitation is that while the CDC reported counts of first-dose vaccination are the best available baseline available for comparison between survey and model estimates and the ``true'' prevalence of first-dose vaccination, these data are imperfect. As we demonstrated in Figure \ref{fig:1b}, the counts for both males and females ages 65 years and older are greater than the total population in these sex and age groups reported in the ACS 5-year estimates.

 \section{Discussion} \label{sec4}
MRP is a quick, computationally inexpensive, and relatively simple-to-implement estimation method. Its ease and flexibility has made it increasingly popular. New advancements in statistical computation now allow us to implement both classic and spatial MRP pipelines for extremely complex Bayesian hierarchical models using INLA. We can now produce and visualize these estimates in minutes. From an implementation standpoint, this is extremely exciting. 

As with any statistical method, MRP has limitations. In this analysis, we were unable to capture the ``true" county-level estimates of first-dose vaccination provided by the CDC for the vast majority of counties, though all four models contained state-level estimates produced by the mean survey weights. Unfortunately, MRP is a method that is only effective when its model accurately reflects the outcome of interest in the target population \citep{KennedyGelman2021}. It will produce estimates for virtually all demographic strata across each geographic unit; it is up to us to validate these. In modelling situations, we make decisions based on expertise provided in the appropriate literature and also according to variables present in the available data. While MRP is able to make predictions for all observed and unobserved areas represented in the survey data used to train the model, the model will be unable to effectively capture the outcome of interest if there is inadequate population and contextual data available with which to predict and poststratify. 

Here, we followed the classic MRP approach of only including covariates present in the data; these were limited because they had to match categorization definitions across data sources (i.e., same age ranges used for age bins, etc.). As previously noted, it has been suggested that COVID-19 vaccine hesitancy in California (and the US, more broadly) had been driven by lower likelihood of vaccination among members of groups historically marginalized according to race/ethnicity \citep[e.g.,][]{NguyenEtAl2022,KhubchandaniMacias2021,WillisEtAl2021}, but by early 2021 COVID-19 vaccine trends in California were significantly driven by and positively associated with greater educational attainment \citep{ThomasEtAl2021}. As the CDC did not release a cross-tabulation of county vaccination counts by sex, age, race/ethnicity, and education, we could not predict the probability of these strata and reliably verify them. Modelling choices in MRP are necessarily based on strong assumptions with respect to demographic structures that underlie the outcome of interest in the target population. Excluding key structuring demographics from the model due to a lack of availability or granularity in the poststratification data suffuses the model with unintended but strong, perhaps even incorrect, assumptions. This was the case with this evaluation: first-dose vaccination estimates are likely to be strongly structured according to sex, age, educational attainment, but this analysis provides evidence that the demographics-specific outcome data represented in this survey sample do not consistently capture important subpopulation vaccination trends. These models were also limited because CDC estimates were only available for very few age categories (18-24 years, 26-64 years, and 65+ years).

Naturally, the question remains: how could we improve these models, and MRP estimation broadly? \citet{KastellecEtAl2015} suggest expanding the poststratification table by using auxiliary survey data including area-level ecological/population characteristics with the aim of reducing the differences between the survey sample and target population. In our case, including race/ethnicity population proportions may improve vaccination estimates, but it remains that the validation data do not contain these details. As \citet{KennedyGelman2021} remind us, we must know our populations and models to produce generalizable results beyond our samples. To this end, another area of improvement is to try to ameliorate the survey sample selection process. As the CTIS was administered through the Facebook platform, user activity drove sample selection, where more frequently active users were more likely to be selected. Therefore, the sample selection process contained an unavoidable temporal process, where in a daily random sample more active users will, of course, be oversampled. In this case, the CTIS significantly oversampled individuals in an over-aggregated age category: aged 25-64 years. The problems inherent to this aggregation decision are difficult to overcome with MRP. It would also be beneficial to be provided better data with respect to survey weights. Companies like Facebook are producing large-scale datasets with relatively principled sampling schemes, but they do not provide the core marginals to accurately adjust the survey data they share. We encourage these companies to share their user-representative metrics so we can benchmark estimates to them.

Clearly, there are important limitations to using MRP in the current health and demographics data landscape. Estimation methods are used when directly observed data are unavailable, such as in emergency response and planning situations. This includes public health agencies' needs with respect to tracking effects and aspects of COVID-19 response during the first years of the pandemic. Estimation methods therefore have the potential to meet a critical need in times of crisis. Better demographics must be recorded and released to researchers to improve local- and state-level response and policy decisions. We must not only work to balance individual privacy concerns with data granularity, but align demographic strata and geographic data choices in survey products with generalizable population data used in national census products, and also make them available to researchers. 

\section{Conclusion}\label{sec5}

MRP is an increasingly popular tool used to adjust non-representative survey data to reflect the target population \citep{GaoEtAl2021}. Many population health outcomes of interest, including vaccination uptake, are known to have both demographic and spatial structure. Both classic and spatial MRP can now be implemented within a Bayesian framework and in a computationally efficient manner using INLA, avoiding MCMC. In this analysis, though model selection preferred both classic and spatial MRP models, these were unable to produce reliable estimates of first-dose vaccination in California for June 30, 2021, in comparison to CDC reported estimates of the same. Data aggregation choices on the part of the CTIS and Facebook are partially the issue. Clearly, further research is necessary to delineate the utility of spatial MRP. We suggest analyses using MRP be undertaken with care, and must use appropriate validation such as would be possible if, for instance in this case, user-representative metrics were available to researchers.

%the JSSR-A format suggests these go after appendices but that makes it more difficult to edit, so I've moved these here for now:
\section{Competing interests}
No competing interest is declared.

\section{Author contributions statement}
%Must include all authors, identified by initials, for example:
A.S. and J.W. conceived the experiments. A.S. and Z.W.A. gained access to the data. A.S. cleaned the data, coded and implemented the experiments, completed analysis, and visualized the results. A.S. wrote the manuscript. A.S., Z.W.A., and J.W. reviewed the manuscript.

\section{Acknowledgments}
Partial support for this research came from a Shanahan Endowment Fellowship and a Eunice Kennedy Shriver National Institute of Child Health and Human Development training grant, T32 HD101442-01, and research infrastructure grant, P2C HD042828, \textit{to the Center for Studies in Demography \& Ecology }at the \textit{University of Washington.} Additionally, partial support came from NSF Grant \#BCS-2028160, and from ARO Award \#W911NF-19-1-0407. This research is based on survey results from Carnegie Mellon University’s Delphi Group. We thank the Delphi Group for access to their data, and Facebook's Data for Social Good Team for access to the accompanying user data and weights.

\bibliographystyle{abbrvnat} %suggested style by JRSS-A
\bibliography{main_bib}

%%%%%%%%%%%%%%
\clearpage
\begin{appendices}

\section{Appendix}
%first table

{\footnotesize
\begin{longtable}{cccccc}
        \caption{Mean survey estimates of first-dose vaccination status from the COVID-19 Trends and Symptoms (CTIS) Survey for June 30, 2021, by age, sex, and educational attainment} \label{tab:ctis_direct_agesexedu} \\
            \toprule
            Vaccination Status & Age & Sex & Education Level & Mean & 95\% Confidence Interval \\ 
            \midrule
            \endfirsthead
            
            Vaccination Status & Age & Sex & Education Level & Mean & 95\% Confidence Interval \\ 
            \midrule
            \endhead
            
            \midrule
            \multicolumn{6}{r}{\textit{Continued on next page}} \\
            \bottomrule
            \endfoot
            
            \bottomrule
            \endlastfoot
            
            No & 18-24 years       & Female & Less than high school                 & 0.069 & 0.029, 0.109 \\
            No & 18-24 years       & Female & High school or equivalent             & 0.394 & 0.318, 0.469 \\
            No & 18-24 years       & Female & Some college                          & 0.374 & 0.300, 0.448 \\
            No & 18-24 years       & Female & Associate's degree                    & 0.078 & 0.037, 0.119 \\
            No & 18-24 years       & Female & Bachelor's degree                     & 0.068 & 0.029, 0.107 \\
            No & 18-24 years       & Female & Professional or graduate degree       & 0.018 & -0.003, 0.038 \\
            No & 18-24 years       & Male   & Less than high school                 & 0.093 & 0.032, 0.154 \\
            No & 18-24 years       & Male   & High school or equivalent             & 0.464 & 0.350, 0.578 \\
\end{longtable}
}

%CDC state counts table
\begin{table}[H]
    \centering
    \footnotesize
    \caption{Centers for Disease Control (CDC) estimated number of first doses administered in California by June 30, 2021, by sex, and age}
     \label{tab:cdc_counts_bysexage} 
    \begin{tabular}{ccccc}
        \toprule
        Total  & Sex    & Age Group      & Estimate & Proportion of Subpopulation \\ 
        \midrule
        25,296,329                  &        &                &                      &                           \\
         & Female & 18-24 years    & 1,112,672            & 0.037                     \\
          & Female & 25-49 years    & 7,610,234            & 0.250                     \\
          & Female & 65+ years      & 4,480,140            & 0.147                     \\
          & Male   & 18-24 years    & 1,044,329            & 0.034                     \\
          & Male   & 25-49 years    & 7,253,978            & 0.239                     \\
        & Male   & 65+ years      & 3,794,976            & 0.125                     \\ \bottomrule
    \end{tabular}
\end{table}
\vspace{10pt}

%CDC county counts table
{\footnotesize
    \begin{longtable}{cccccc}
        \caption{Centers for Disease Control (CDC) estimated number of first doses administered in California by June 30, 2021, by county} \label{tab:cdc_counts_bysexage_county} \\ 
        \toprule
        County & Administered First Doses & County Population & Proportion with First Dose \\ \midrule
        \endfirsthead
    
       County & Administered First Doses & County Population & Proportion with First Dose \\ \midrule
        \endhead
    
        \midrule
        \multicolumn{5}{r}{\textit{Continued on next page}} \\ 
        \endfoot
    
        \bottomrule
        \endlastfoot
     
        Kern & 328433 & 634299 & 0.518 \\
        Calaveras & 22930 & 38210 & 0.600 \\
        San Mateo & 517120 & 608854 & 0.849 \\
        Alameda & 1075757 & 1319993 & 0.815 \\
        Lassen & 8517 & 25663 & 0.332 \\
        Santa Barbara & 246398 & 345856 & 0.712 \\
        Marin & 184868 & 207467 & 0.891 \\
        Shasta & 70861 & 140599 & 0.504 \\
\end{longtable}
}
\vspace{10pt}

{\footnotesize
\begin{landscape} % Begin landscape mode
    \begin{longtable}{p{1.25cm}p{7cm}ccccccccc}
        \caption{Bayesian hierarchical logistic regression results for all models}
         \label{tab:inla_regression_results} \\
        \toprule
        Type & Model Name  & Sex & LCPO & Mean & SD & 0.025quant & 0.5quant & 0.975quant & KLD \\ 
        \midrule
        \endfirsthead
        
        Type & Model Name  & Sex & LCPO & Mean & SD & 0.025quant &  0.5quant & 0.975quant & KLD \\ 
        \midrule
        \endhead
        
        \midrule
        \multicolumn{11}{r}{\textit{Continued on next page}} \\
        \endfoot
    
        \bottomrule
        \endlastfoot

    Spatial & fixed effect on age, BYM2 by education & female & 0.32 & 0.955 & 0.216 & 0.526 & 0.954 & 1.39 & 1.41E-07 \\
    Spatial & fixed effect on age, BYM2 by education & female & 0.32 & 0.598 & 0.057 & 0.487 & 0.598 & 0.71 & 5.53E-11 \\
    Spatial & fixed effect on age, BYM2 by education & female & 0.32 & -0.224 & 0.193 & -0.614 & -0.223 & 0.161 & 0.00000133 \\
    Spatial & fixed effect on age, BYM2 by education & female & 0.32 & -0.44 & 0.186 & -0.817 & -0.439 & -0.0661 & 0.0000025 \\
    Spatial & fixed effect on age, BYM2 by education & female & 0.32 & -0.294 & 0.185 & -0.669 & -0.294 & 0.0785 & 0.00000276 \\
    Spatial & fixed effect on age, BYM2 by education & female & 0.32 & -0.201 & 0.188 & -0.582 & -0.2 & 0.177 & 0.00000202 \\
    Spatial & fixed effect on age, BYM2 by education & female & 0.32 & 0.464 & 0.187 & 0.0883 & 0.463 & 0.842 & 0.00000234 \\
    Spatial & fixed effect on age, BYM2 by education & female & 0.32 & 0.74 & 0.19 & 0.363 & 0.739 & 1.13 & 0.00000187 \\
\bottomrule
    \end{longtable}
\end{landscape}
}

{\footnotesize
\begin{longtable}{ccp{4.5cm}ccc}
    \caption{Posterior median county-level estimates of the proportion of the population administered with a first-dose vaccination by June 30, 2021 in California, including CDC and mean survey estimates} \label{tab:county_propestimates} \\
    \toprule
    Type & County & Model/Estimation Method  & Estimate & 95\% CI: Lower & 95\% CI: Upper \\ 
    \midrule
    \endfirsthead
    
    \toprule
    Type & County & Model/Estimation Method & Estimate & 95\% CI: Lower & 95\% CI: Upper \\ 
    \midrule
    \endhead
    
    \midrule
    \multicolumn{6}{r}{\textit{Continued on next page}} \\
    \endlastfoot

    \multicolumn{6}{r}{\textit{Continued on next page}} \\
    \endfoot

    \endlastfoot
        
     & Alameda & CDC reported count & 0.815 & - & - \\
     & Alameda & mean survey estimate & 0.931 & 0.917 & 0.943 \\
    Non-Spatial & Alameda & fixed effect on age, RW(1) by education, IID by county & 0.935 & 0.917 & 0.948 \\
    Non-Spatial & Alameda & fixed effect on age, fixed effect on education, IID by county & 0.934 & 0.918 & 0.947 \\
    Spatial & Alameda & fixed effect on age, BYM2 by education & 0.889 & 0.877 & 0.9 \\
    Spatial & Alameda & fixed effect on age, RW(1) by education, BYM2 by county & 0.935 & 0.92 & 0.948 \\
     & Alpine & CDC reported count & 0 & - & - \\
    Non-Spatial & Alpine & fixed effect on age, RW(1) by education, IID by county & 0.886 & 0.73 & 0.961 \\
\bottomrule
\end{longtable}
}

{\footnotesize
\begin{longtable}{cp{7cm}cccc}
        \caption{Posterior median aggregated state-level estimates of the proportion of the population administered with a first-dose vaccination by June 30, 2021 in California} \label{tab:state_propestimates} \\
        \toprule
        Type & Model  & Estimate & 95\% CI: Lower & 95\% CI: Upper \\ 
        \midrule
        \endfirsthead
        
        \toprule
        Type & Model & Estimate & 95\% CI: Lower & 95\% CI: Upper \\ 
        \midrule
        \endhead

    Non-Spatial & fixed effect on age, RW(1) by education, IID by county & 1.76 & 1.7 & 1.8 \\
    Non-Spatial & fixed effect on age, fixed effect on education, IID by county & 1.75 & 1.7 & 1.8 \\
    Spatial & fixed effect on age, BYM2 by education & 1.75 & 1.73 & 1.78 \\
    Spatial & fixed effect on age, RW(1) by education, BYM2 by county & 1.76 & 1.7 & 1.8 \\

    \bottomrule
\end{longtable}
}

\end{appendices}

\end{document}